\documentclass{ws-jai}

\usepackage[dvipsnames]{xcolor}

\begin{document}

\catchline{}{}{}{}{} % Publisher's Area please ignore

\markboth{Young Min Seo}{Statistical Prediction of [CII] Observations}

\title{Statistical Prediction of [CII] Observations by Constructing Probability Density Functions using SOFIA, Herschel, and Spitzer Observations}

\author{Young Min Seo$^{1}$, Karen Willacy$^{1}$, Umaa Rebbapragada$^{1}$}

\address{
$^{1}$Jet Propulsion Laboratory, California Institute of Technology, 4800 Oak Grove Drive, Pasadena, CA, 91109, USA\\
}

\maketitle

\corres{$^{1}$Young Min Seo, youngmin.seo@jpl.nasa.gov}

\begin{history}
\received{(to be inserted by publisher)};
\revised{(to be inserted by publisher)};
\accepted{(to be inserted by publisher)};
\end{history}

\begin{abstract}

We present a statistical algorithm for predicting the [CII] emission from {\it Herschel} and {\it Spitzer} continuum images using probability density functions between the [CII] emission and continuum emission. The [CII] emission at 158 $\mu$m is a critical tracer in studying the life cycle of interstellar medium and galaxy evolution. Unfortunately, its frequency is in the far infrared (FIR), which is opaque through the troposphere and cannot be observed from the ground except for highly red-shifted sources (z $\gtrsim$ 2). Typically [CII] observations of closer regions have been carried out using suborbital or space observatories. Given the high cost of these facilities and limited time availability, it is important to have highly efficient observations/operations in terms of maximizing science returns. This requires accurate prediction of the strength of emission lines and, therefore, the time required for their observation. However, [CII] emission has been hard to predict due to a lack of strong correlations with other observables. Here we adopt a new approach to making accurate predictions of [CII] emission by relating this emission simultaneously to several tracers of dust emission in the same region. This is done using a statistical methodology utilizing probability density functions (PDFs) among [CII] emission and {\it Spitzer} IRAC and {\it Herschel} PACS/SPIRE images. Our test result toward a star-forming region, RCW 120, demonstrates that our methodology delivers high-quality predictions with less than 30\% uncertainties over 80\% of the entire observation area, which is more than sufficient to test observation feasibility and maximize science return. The {\it pickle} dump files storing the PDFs and trained neural network module are accessible upon request and will support future far-infrared missions, for example, GUSTO and FIR Probe.     

\end{abstract}


\section{Introduction} \label{sec:intro}

Making accurate predictions of proposed scientific measurements is crucial for planning space missions and writing successful research proposals. In astronomy, a prediction is typically a simulated observable (e.g., images or spectra) of the target object and is a critical requirement for justifying feasibility in observing proposals because a small decrease in the flux of the target object results in a large increase in observation time (signal-to-noise $\propto$ $\sqrt{\rm time}$). However, predicting observables has typically been challenging since it frequently involves heavy numerical modeling/simulations of target objects, which consume significant resources. The theoretical modeling can be avoided only when there is a simple and tight correlation between the existing and target observables, for example, $^{12}$CO 1-0 and dust continuum flux. Unfortunately, many physical and chemical processes in astronomy are highly complex, resulting in no simple correlation. A methodology that can accurately predict observables based on readily available existing data be a convenient and cost-effective alternative to intensive numerical modeling for every source.

The strength of [CII] emission, which is the emission coming from singly ionized carbon, C$^+$, is difficult to predict because it lacks a tight correlation to other observables. [CII] 158 $\mu$m emission is a key emission line for probing star formation in the Milky Way, galaxy evolution, and planet-forming disk evolution since it is one of the main coolants of the heated gas tracing UV irradiated regions \citep[e.g.,][]{pineda13, langer14}. Although of high scientific importance, observing [CII] is challenging because our atmosphere is opaque at 158 $\mu$m, and [CII] can only be observed above the stratosphere or in space, leading to high costs and significant effort in observations. Therefore, accurate prediction of [CII] observations is critical to maximizing the efficiency of observations and science return. 

Predictions of [CII] emission have been made by intensive numerical simulations \citep[e.g.,][]{petit06,decataldo17,ferrara19}. CO or [CI] emissions also have been used to predict [CII] emission, but the correlation is weak, resulting in large uncertainties in the predicted [CII] emission flux. The objects targeted for [CII] observations (e.g., star-forming regions, galaxies) have often been well-observed in infrared (IR) and sub-mm wavelengths using {\it Herschel} and {\it Spitzer}, resulting in a large archival database covering a range of wavelengths. Here we present a new methodology for accurately predicting [CII] line emission based on the {\it Herschel} and {\it Spitzer} archival data, using probability density functions and neural networks. The method is highly relevant to future [CII] observations and IR space/suborbital mission planning and has general applications beyond the [CII] example.

In \S2 we briefly overview the new methodology and the applications to the prediction of [CII] emission from archival images of {\it Herschel} and {\it Spitzer}. In \S3, we elaborate on the prediction quality compared to the observations. Finally, we discuss and summarize the results in \S4.

\section{Data Processing Steps and Algorithms} \label{sec:algorithm}

The problem addressed in this work may be simplified as follows: we have many different sets of input data, but each of them is only loosely correlated to the output data that we want to predict. Thus, a single set of input data may predict the output data with significant uncertainty. On the other hand, the output data may be accurately determined if multiple input datasets are used simultaneously, constraining the output uncertainty. The key approach in this study is that we evaluate statistical correlations of [CII] to other observations using probability density functions (PDFs), from which we can predict the most probable [CII] images with minimal uncertainty. This section explains the overall steps and algorithms, including the data required to build the PDFs, the necessary preprocessing of the input data, the steps taken to evaluate the PDFs, and the prediction methodology using PDFs.

\subsection{{\it Herschel}, {\it Spitzer}, and SOFIA data} \label{sec:data}

We use {\it Herschel} images, including dust continuum at 70 $\mu$m, 160 $\mu$m, 250 $\mu$m, 350 $\mu$m, and 500 $\mu$m, and {\it Spitzer} observations at 3.6, 4.5, 5.8, and 8 $\mu$m, which includes emissions from polycyclic aromatic hydrocarbons (PAHs) at 3.6 and 8 $\mu$m, for the input data. We focus on {\it Herschel} and {\it Spitzer} observations as input datasets for predicting the [CII] intensity because the two surveys are among the highest quality surveys having the most extensive coverage of the sky. If [CII] intensity can be accurately predicted using two surveys only, then the method proposed here could be used to [CII] in a large number of star-forming regions. On the other hand, using these two input datasets might not trace all of the physical and chemical processes related to [CII] emission since the two IRAC bands of {\it Spitzer} observations (3.6 and 8 $\mu$m) are tied to PAH emissions, and the five PACS/SPIRE bands of {\it Herschel} observations are correlated to each other by the dust temperature. We found that we can predict [CII] emissions using only the two surveys, which we will further elaborate on in the next sections. There are other observations related to [CII] (e.g., high-$J$ CO) at a comparable angular resolution to the {\it Herschel} and {\it Spitzer data} that may diversify the input data set. However, the sky coverage is very small, which would limit the prediction of [CII] emission to a small number of star-forming regions. Thus, we limit our analysis and application to the statistical relationship between the continuum emissions of {\it Herschel} and {\it Spitzer} and the [CII] emissions.

We use SOFIA [CII] data downloaded from the NASA/IPAC Infrared Science Archive (IRSA). SOFIA [CII] observations are the largest [CII] surveys with an angular resolution comparable to the {\it Herschel} and {\it Spitzer} observations. Still, the SOFIA [CII] surveys cover less than 1\% of regions surveyed by {\it Herschel} and {\it Spitzer}; therefore, the majority of the sky is not observed in [CII] and extending coverage of this ion is a major goal of future FIR telescopes (e.g., GUSTO, \citet{goldsmith22} and FIR probe concepts). 

To make the proper statistical analysis, the quality of datasets needs to be comparable. The {\it Herschel} and {\it Spitzer} data \citep{Churchwell09,molinari16} are two-dimensional images taken with highly sensitive direct detectors. SOFIA [CII] observations are spectral-image data observed using a heterodyne receiver (GREAT, \citet{risacher18}). While the SOFIA GREAT instrument is the most sensitive far-IR heterodyne receiver, and its data quality is one of the best in the community, the noise level of the SOFIA images is typically higher than that of {\it Herschel} and {\it Spitzer} images. To evaluate the PDFs, the SOFIA data noise level needs to be as low as possible or at least comparable to {\it Herschel} and {\it Spitzer} images. We use a noise reduction convolutional neural network (CNN) \citep[e.g.,][, also see Appendix A]{seo20}, which typically delivers a factor of a few to an order of magnitude reduction in the noise. We found that using this approach the noise level of the SOFIA [CII] data is typically comparable to, or slightly higher than that of {\it Herschel} and {\it Spitzer} images. 

The {\it Herschel} and {\it Spitzer} images are downloaded from Herschel Science Archive and IRSA, respectively, and used without any further processing. While the data have been processed with a generic pipeline, the noise and quality of the data are considerably better than SOFIA data in general. Also, we did not correct any foreground or background continuum emissions. Conventionally, in deriving physical parameters of star-forming regions using the dust continuum (e.g., hydrogen column density), the dust continuum from local foreground or background clouds along the lines of sight should be subtracted to obtain only the emissions from the star-forming region \citep[e.g.,][]{guzman15}. In this study, the main methodology is based on the statistics of large-scale samplings in multiple star-forming regions. The main statistical characteristics are dominated by global properties between the continuum and [CII] emissions, while the local foreground or background clouds may be expressed as noise/uncertainty of the global statistics (e.g., increasing scattering of data distribution in a scatter plot). Thus, unless foreground or background clouds span a significant fraction of the star-forming region, the impact of foreground or background clouds on the [CII] prediction would be negligible; thus, we use the continuum data without any additional processing.

For our proposed method to apply to diverse star-forming regions, we need data from regions with a wide range of physical and chemical conditions. We select three star-forming regions, M16, M17, and NGC 6334, which are all observed in SOFIA (FEEDBACK, \citealt{schneider20}), {\it Herschel} \citep{molinari16}, and {\it Spitzer} \citep{Churchwell09}. We also use RCW 120, another star-forming region, to validate our methodology. We chose these star-forming regions for two reasons. First, these star-forming regions are relatively close to us ($\leq$ 2 kpc), giving rich details of the star-forming processes and low-noise images. Second, they cover a wide range of the astrophysical and chemical properties found in the ISM. They contain high-mass OB stars ionizing the surrounding molecular clouds and low-mass stars in the molecular clouds. The molecular clouds are exposed to a wide range of UV flux depending on their distance to the high-mass stars, experiencing highly diverse physical and chemical processes (G$_0$ = 0 $-$ 10$^6$, where G$_0$ is local UV radiation flux. 1 G$_0$ = 5.29 $\times$ 10$^{-14}$ erg/s, \citealt{mathis83}). In addition, M17 had a recent supernova and is exposed to strong X-ray \citep{townsley03}, inducing extreme physical and chemical processes. Other star-forming regions are observed in [CII] using SOFIA, but the ones used here have the lowest noise and include most of the physical and chemical conditions that could be found in other star-forming regions.

\begin{figure*}[tb]
\centering
\includegraphics[angle=0,scale=0.45]{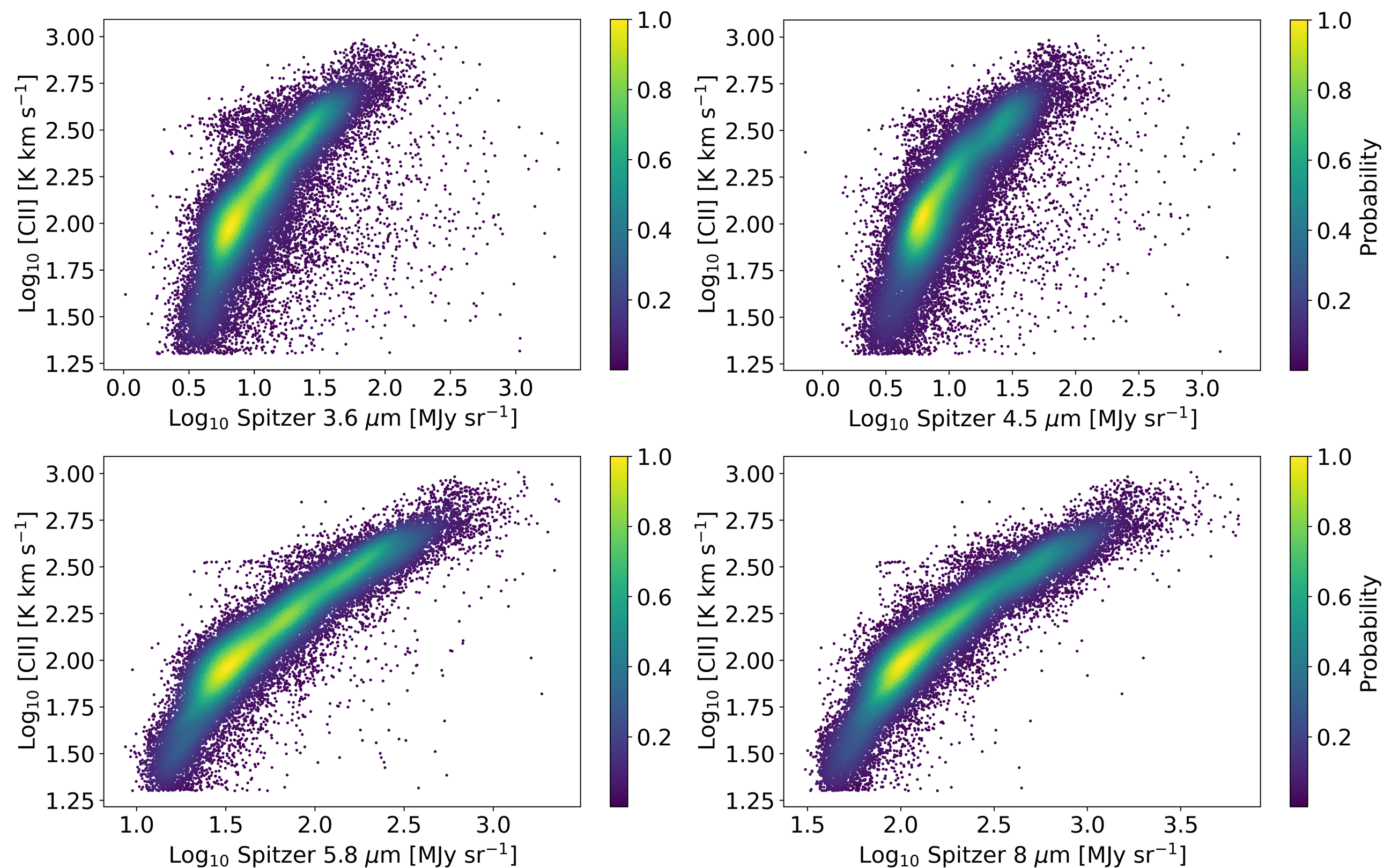}
\caption{Statistical relationships between intensities of [CII] and the continuum at 3.6, 4.5, 5.8, 8, 70, 160, 250, 350, and 500 $\mu$m in three star-forming regions -- M16, M17, and NGC 6334. The dust continuum at 3.6, 4.5, 5.8, and 8 $\mu$m data is the {\it Spitzer} IRAC observations. The dust continuum data at 160, 250, 350, and 500 $\mu$m are the {\it Herschel} PACS \& SPIRES observations. Each scatter point corresponds to each pixel in the SOFIA [CII] images. The color bar indicates the relative probability density estimated using Gaussian kernel density estimation (KDE) in two-dimensional space. Distributions of the data do not show significantly tight correlations between [CII] and the continuum but are rather widely scattered. The continuum emissions at 3.6 and 8 $\mu$m, which are related to PAHs, tend to have tighter relationships with the [CII] intensity compared to the dust continuum at longer wavelengths ($>$ 200 $\mu$m). Continued to Fig. \ref{f2}.}
\label{f1}
\end{figure*}

\begin{figure*}[tb]
\centering
\includegraphics[angle=0,scale=0.65]{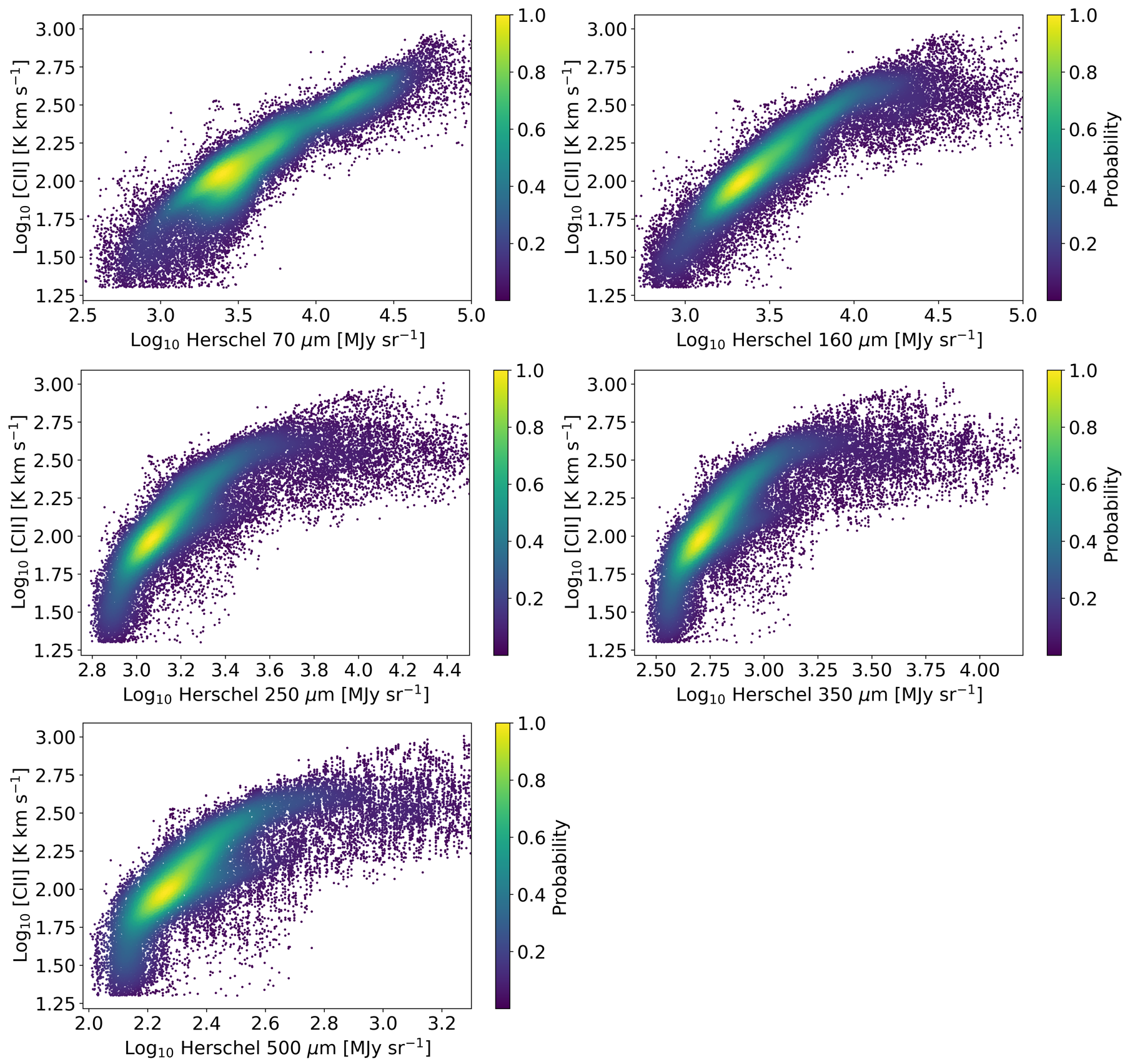}
\caption{continued from Fig. \ref{f1}. }
\label{f2}
\end{figure*}

\subsection{Building probability density functions} \label{sec:pdf}

We evaluate the probability density functions (PDFs) representing the statistical relationship between the [CII] and continuum emission at each wavelength. Figures \ref{f1} \& \ref{f2} show the pixel-by-pixel comparison between the {\it Herschel} and {\it Spitzer} images and the SOFIA integrated [CII] map in M16, M17, and NGC 6334. The data points are widely scattered and show weak correlations, particularly between [CII] and the dust continuum at $>$200 $\mu$m. The [CII] and {\it Spitzer} IRAC observations show relatively tighter correlations, similar to the results from other [CII] studies \citep[e.g.,][]{pabst21}. A tighter correlation at a mid-IR wavelength rather than a far-IR wavelength is expected since the [CII] emission comes from a region heated by UV, and the 8 $\mu$m emission is mostly from UV-heated PAHs. In contrast, the 350 $\mu$m emission is from dust grains in cold regions, where carbon ionization is minimal. Within {\it Spitzer} IRAC observations, 8 $\mu$m emissions show the tightest correlation to the [CII] emission while the overall trend is similar. The 3.6 $\mu$m band also includes PAHs emission, so the 3.6 and 8 $\mu$m emissions are correlated through the radiative transfer processes of PAHs, but there are significant photons coming from stars at $<$8 $\mu$m, which increases scatter in the Figure \ref{f1}. 

However, even the scatter in the 8 $\mu$m emission is too large to be used alone to predict [CII] emission. Also, the distribution of data points, particularly between [CII] and 350 $\mu$m, shows a complicated (multimodal) distribution, which may originate from the highly complicated physical and chemical processes along the line of sight toward the three star-forming regions. For example, the dust continuum at 350 $\mu$m may trace multiple dense clouds along the line of sight, while the [CII] emission is mainly from the irradiated warm surface of the dense clouds and ionized regions. Thus, it is practically impossible to construct a parametric model (parametric inference) for the distributions. On the other hand, we may estimate the probability of the [CII] intensity as a function of the continuum intensity from the data in Figures \ref{f1} \& \ref{f2} to understand the relationships between [CII] and continuum emission, which is a classical/frequentist statistical approach of evaluating a probability \citep[e.g.,][]{rosenblatt56,parzen62}. The color of scatter dots in Figures \ref{f1} and \ref{f2} show the two-dimensional PDFs using the Gaussian kernel density estimation (KDE) in the Python {\it scikit-learn} package, which is a non-parametric kernel density estimation \citep[e.g.,][]{jones96}. The Gaussian bandwidth/kernel size is automatically optimized to minimize the noise. 

\begin{figure*}[tb]
\centering
\includegraphics[angle=0,scale=0.47]{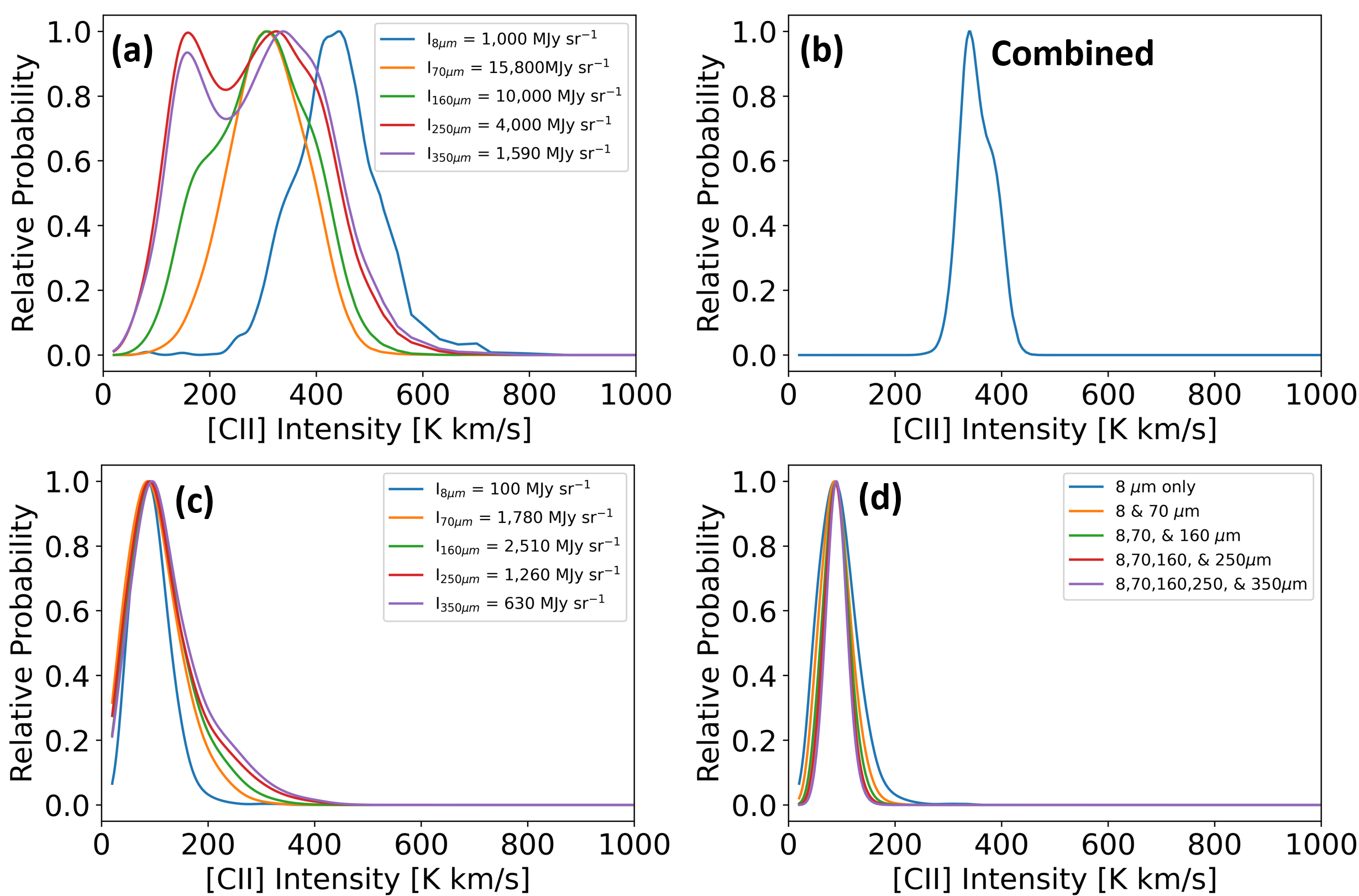}
\caption{(a) The individual probability density function (PDF) of [CII] intensity for given dust continuum intensities at five different wavelengths. The intensities of the five continuum emission are written in the plot. The PDFs are estimated from the data points in Figure \ref{f1} using the Gaussian kernel density estimation. (b) The combined PDF of [CII] intensity when the five continuum intensities constrain the [CII] intensity. The combined PDF is obtained by multiplying the eight PDFs in panel (a). (c) Another example of PDFs with a different set of continuum intensities. (d) Results predicted by combining different numbers of PDFs The width of the combined PDF decreases by 40 \% from the PDF of 8 $\mu$m only to the combined PDF of five different observations.} 
\label{f3}
\end{figure*}

Figure \ref{f3} (a) shows the relative probability density functions (PDFs) when the intensities at 8, 70, 160, 250, 350, and 500 $\mu$m are given (or observationally determined). We omitted the PDFs of 3.6, 4.5, and 5.8 $\mu$m in Figure \ref{f3} for readability since these three PDFs are practically identical to the PDF of 8 $\mu$m although slightly broader. The highest peak of the PDF indicates the most probable [CII] emission for a given intensity of the continuum emission. As the figures show, the PDFs may have multiple peaks, demonstrating that constraining the [CII] emission using a single continuum emission may have large uncertainty. For example, the PDF of [CII] and 350 $\mu$m shows that the most probable [CII] intensity will be around 325 K km/s when we observed 350 $\mu$m continuum intensity of 1,590 MJy/sr. But there is another peak in the probability density near 133 K km/s. The 1-$\sigma$ uncertainty of [CII] emission (68\% area coverage below the density function with the density peak as the center) covers both peaks and is as wide as 360 K km/s. Such large uncertainty, a factor of two in intensity, is inadequate for testing observation feasibility since it results in a factor of four uncertainty in observation time. A similar shape (e.g., multiple peaks) is also found in the PDFs of the 8, 250, and 350 $\mu$m images, while the PDFs of 70 and 160 $\mu$m show single-peak functions. The 1-$\sigma$ uncertainty of the [CII] emission using the 8 $\mu$m data is a factor of a few smaller than that using the 350 $\mu$m data at the given intensity. However, the 1-$\sigma$ uncertainty of the [CII] emission is not constant over the 8 $\mu$m intensity and is much higher at a higher 8 $\mu$m intensity. Such an increase in the uncertainty of [CII] emission predicted using 8 $\mu$m data confirms the difficulty in predicting the [CII] emission using a {\it single} observation. 

To improve the accuracy of the predicted [CII] emission and reduce the 1-$\sigma$ uncertainty, we combine the eight PDFs at 3.6, 4.5, 5.8, 8, 70, 160, 250, and 350 $\mu$m. The 500 $\mu$m emission is neglected in this step since it is similar to the 350 $\mu$m data, and the PDF of 500 $\mu$m is the broadest of the PDFs; hence it would only provide a marginal constraint to the [CII] intensity and its uncertainty. To combine the eight PDFs, they are multiplied without any weighting, which is equivalent to equal weighting. The weighting factor of PDF denotes the dependency among observations. If each PDF is estimated from independent observations, the weighting factors should be all unity. On the other hand, if a prior observation influences the results of the following observations, the following observations become conditional events. In such a case, weight should be determined to reflect the conditional probability. In this study, the intensities at 3.6, 4.5, 5.8, 8, 70, 160, 250, and 350 are independently measured values (independent events), and we combine the PDFs by multiplying them without any weighting.

Figure \ref{f3} (b) shows the PDF of the [CII] intensity constrained with eight intensities, which is obtained by multiplying the eight PDFs in Figure \ref{f3} (a). It shows a single-peaked density profile with its 1-$\sigma$ width being $\pm$40 K km/s approximately, corresponding to roughly a 12\% error to the peak value and hence to only a 20\% uncertainty in the observation time, which is adequate to be used for planning scientific observations. Thus, we may make a prediction of the [CII] image of a star-forming region if the region has been observed using {\it Herschel} and {\it Spitzer}. 

To understand the uncertainty of the predicted [CII] intensity depending on the selection of the PDFs, we present another example of the PDFs with a different intensity set of the {\it Herschel} and {\it Spitzer} observations in Figure \ref{f3} (c) and (d). Here, the PDFs have similar shapes but different widths. Figure \ref{f3} (d) shows the combined PDFs with a different number of PDFs. As the number of PDFs that are combined increases, the width of the combined PDF decreases, reducing the uncertainty in the predicted [CII] intensity. In particular, combining the narrowest PDFs (PDFs of 8 and 70 $\mu$m) decreases the uncertainty significantly, while broader PDFs reduce the uncertainty only marginally. Compared to using the 8 $\mu$m PDF only, we found that combining the five PDFs results in a 1-$\sigma$ width that is narrower by 40 \%, while the most probable values of [CII] are practically the same. This demonstrates that the methodology also works with two or three PDFs (e.g., choosing the first two or three narrowest PDFs). However, the width of PDFs varies with different ranges of continuum intensities. The PDFs of 8 $\mu$m and 70 $\mu$ are the narrowest two PDFs in general but not always throughout the entire range of the [CII] intensity. Thus, combining as many PDFs as possible will result in the lowest uncertainty of the [CII] prediction.  

\begin{figure*}[tb]
\centering
\includegraphics[angle=0,scale=0.9]{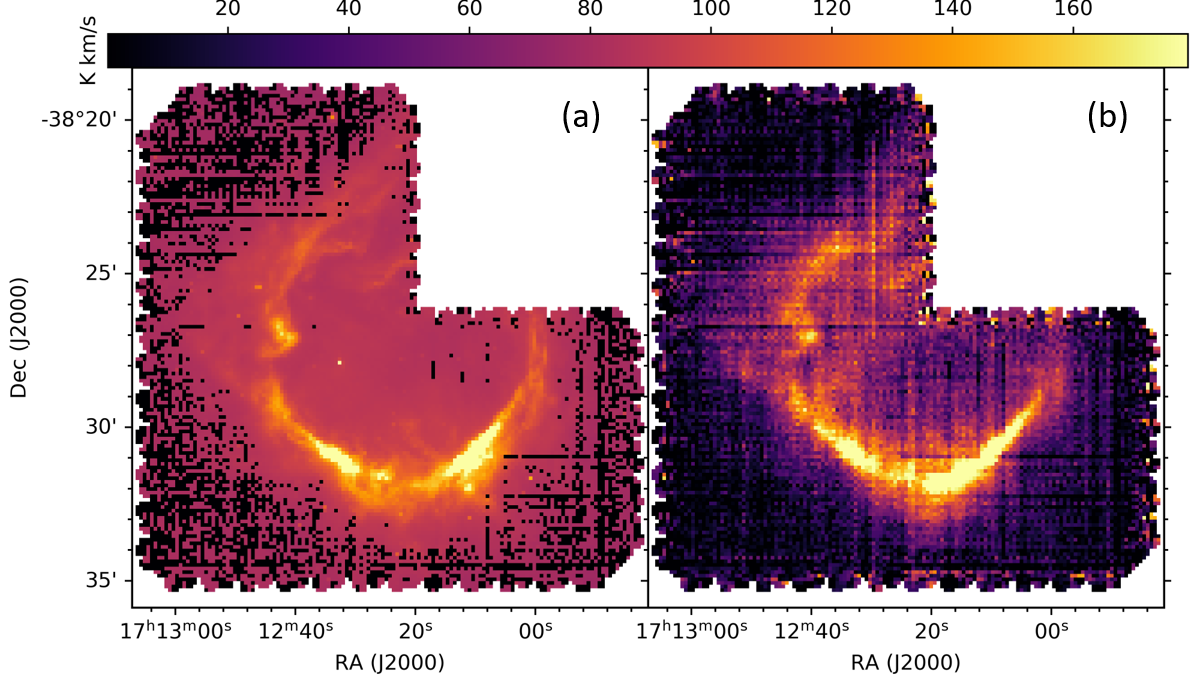}
\caption{The example of [CII] intensity prediction using our algorithm toward the RCW 120 star-forming region. The left panel shows the prediction image, while the right panel is the actual observation using the SOFIA GREAT instrument. The discrepancy between the two images is sufficiently small ($<$50\% to the observed intensities in most of the area), demonstrating our algorithm can predict the [CII] intensity accurately. }
\label{f4}
\end{figure*}

\subsection{Neural network for versatility} \label{sec:NN}

A prediction of the [CII] image may be obtained by using the two-dimensional PDFs in Figures \ref{f1} \& \ref{f2}, evaluating one-dimensional PDFs for given intensities at the eight wavelengths (Figure \ref{f3}), and multiplying the PDFs. We saved the two-dimensional PDFs between [CII] and the continuum emissions using {\it pickle} module in Python. The saved pickle dump files are publicly available on the author's GitHub. The {\it pickle} dump file can be imported by the same version of {\it pickle} module, (3.10.6), and can be used to produce the most probable [CII] images. However, we found that repeatedly calling the two-dimensional PDFs and evaluating one-dimensional PDFs to get the most probable [CII] image is significantly time-consuming for a moderate size of images ($>$10,000 pixels). Also, the {\it pickle} module often changes file structures as the version is updated, so one needs to keep the same version of {\it pickle} module to read in the PDFs correctly, which is troublesome when other Python modules require the latest version of {\it pickle}. Thus, we have developed a more efficient way to produce the [CII] prediction by adopting a neural network.         

Simple neural networks may work as a global function regressor/fitter. The problem of predicting [CII] using continuum emission at five wavelengths is fundamentally the same as evaluating a function that takes five inputs (the intensities of the 8, 70, 160, 250, and 350 $\mu$m emissions) and delivers one output (the [CII] intensity). We use only five inputs since we found that eight inputs take too long to train the neural network. We decided to omit the 3.6, 4.5, and 5.8 $\mu$m data because they have fewer useable data points due to a considerable number of pixels in the images being saturated by bright stars. Therefore, we may generate a large set of the five inputs (continuum intensities) and the output (the most probable [CII] emission) and train the neural network to regress the function for evaluating the five intensities to deliver the most probable [CII] intensity. From the two-dimensional PDFs in Figures \ref{f1} and \ref{f2}, we wrote a data table containing 100 million sets of the five input intensities and the corresponding most probable [CII] intensity for training the neural network and another 10 million sets for validating the neural network. Using such a large table, we trained a simple neural network to estimate the most probable [CII] intensity when intensities at the five wavelengths are given as an input data set. The neural network is a simple deep feed-forward network with three hidden layers (15, 21, and 15 neurons) and has neurons fully connected. The numbers of the hidden layers and the neurons at each layer are adjusted to minimize the computation cost while maximizing the accuracy of the output through trials and errors. {\it Adam} optimizer and the loss function of mean absolute error are used for training the neural network. The training of a neural network with a table of 100 million rows and the validation of the training using a table with 10 million rows took roughly 14 hours using two 2.6 GHz Xeon CPUs having 40 cores in total and a single run of the trained neural network takes less than 0.1 seconds. The loss function converged after training $\sim$7 million rows. The fully trained neural network with 100 million rows shows a typical error of 10\% to the true solution.

\section{Prediction, Uncertainty, and Further Applications} \label{sec:pred}

Figure \ref{f4} compares the [CII] images of the RCW 120 star-forming region that are predicted using the neural network and the actual observations. The two images are very similar down to small [CII] structures in the image, demonstrating that our algorithm accurately predicts the [CII] observation from {\it Herschel} and {\it Spitzer} continuum images. The median error of the predicted image to the observed image is 50\% for the area with [CII] intensity above 25 K km/s, which is roughly 70\% of the entire image, and 24\% for the area with [CII] intensity above 50 K km/s, which is roughly 40\% of the image. The error of prediction is smaller for the brighter [CII] regions since the signal-to-noise level of the data used in the algorithm is significantly better in these regions in the {\it Herschel} and {\it Spitzer} continuum images. These error values are comparable to the calibration uncertainty in [CII] observations ($\sim$30 $-$ 50\%). They are sufficiently small for the predicted images to be used to check observation feasibility and to configure instruments for future [CII] observations.

\begin{figure*}[tb]
\centering
\includegraphics[angle=0,scale=0.5]{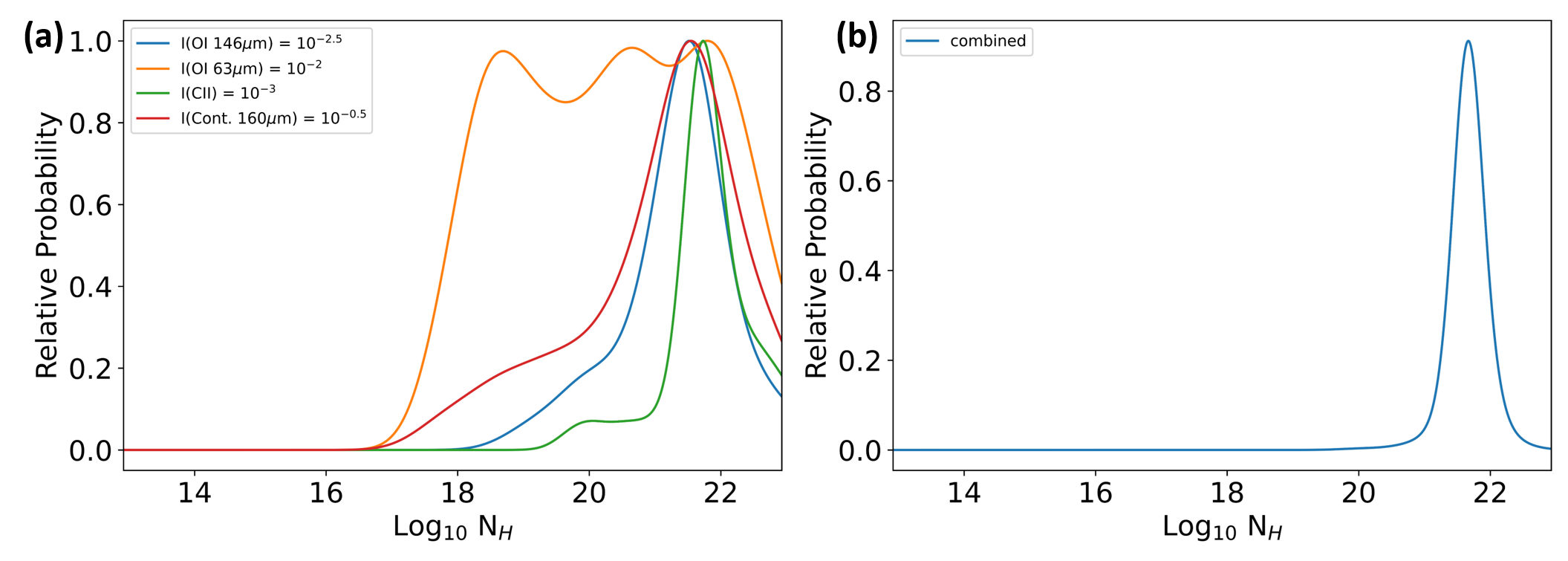}
\caption{(a) The individual probability density functions (PDF) of the hydrogen column density for given [CII], [OI], and dust continuum intensities. (b) The combined PDF of the hydrogen column density when [CII], [OI], and dust continuum intensities constrain the hydrogen column density. The combined PDF is evaluated by multiplying the PDFs in panel (a).} 
\label{f5}
\end{figure*}

While this study has focused on predicting [CII] observations, we can expand the concept beyond [CII] predictions as long as there is a large database to evaluate PDFs. Figure \ref{f5} (a) shows such examples, which are the PDFs between the hydrogen column density, N$_H$, and the observables, including [CII] emission, [OI] emission at 63 $\mu$m and 146 $\mu$m, and dust continuum emission at 160 $\mu$m. The data for evaluating the PDFs are the Monte Carlo models created using the Meudon PDR code \citep{petit06}. The Meudon PDR code simulates physical structures and observables in photodissociation regions (PDRs) of star-forming regions and is one of the most advanced PDR codes in the community. With the PDFs, we can statistically determine the most probable physical parameters from the observables and quantify the uncertainties in the parameters. Figure \ref{f5} (b) shows the PDF of the hydrogen column density, which is obtained by multiplying the four PDFs in Figure \ref{f5} (a). The PDF shows the most probable hydrogen column density is N$_H$ = 10$^{21.7 \pm 0.4}$ cm$^{-2}$. This demonstrates that our methodology can be applied to estimate physical parameters from observables with uncertainty quantification.   

\section{Summary} \label{sec:summary}

[CII] observations are one of the key goals of multiple NASA missions, including SOFIA, GUSTO, and future FIR missions, because it is closely related to star formation and galaxy evolution. To date, the entire [CII] observations in the community cover only a tiny fraction ($<$1\%) of the {\it Herschel} and {\it Spitzer} continuum surveys. Thus, a lack of [CII] measurements has been a major obstacle in the studies of star formation and galaxy evolution. On the other hand, [CII] observations are expensive since they can be done only above the Earth's troposphere, either in the stratosphere or in space. Thus, high operation efficiency is critical to a successful mission and maximizing science returns. Knowing the feasibility of the observation through an accurate prediction of the line strength is the first step in maximizing efficiency. We have thus written an algorithm to make an accurate prediction of [CII] emission from {\it Herschel} and {\it Spitzer} images. The key efforts and results of this study are as follows:  

1. The continuum emission at 3.6, 4.5, 5.8, 8, 70, 160, 250, 350, and 500 $\mu$m have correlations to the [CII] emission. However, continuum emission at a single wavelength does not have a tight correlation sufficient to evaluate [CII] emission within a reasonable uncertainty. This confirms that a parametric inference (e.g., polynomial curve) is insufficient to correlate between continuum emission and [CII], which makes it hard to predict the [CII] emission using a simple curve.  

2. We have demonstrated that using PDFs (non-parametric statistical inference) makes an accurate prediction of the [CII] emission from the continuum emission. The statistical inference may facilitate combining multiple input images to constrain the target observable when the relationship between the input images and the target observable is rather complicated. Our results show that the [CII] emission can be constrained with $\sim$10\% uncertainty in its prediction and $\sim$25\% uncertainty in the prediction of the observation time.       

3. The comparison between the [CII] prediction of RCW 120 using the algorithm and the actual observation shows that the [CII] prediction is accurate for the region with the bright emission ($>$50 K km/s) with $<$25\% of the error, while the error increases toward weak emission ($\sim$25 K Km/s). The errors are comparable to the calibration uncertainty of typical [CII] observations, and the details of the [CII] prediction adequately match the actual observation. Therefore, the prediction in this study is sufficiently accurate for observation planning.      

Beyond making a prediction of observations, the methodology can be further expanded to scientific analysis. The essence of the idea in the methodology is to use a statistical inference of the multiple inputs to the outputs when the relationships between inputs and outputs are too complicated to be expressed using a simple parametric relationship. A large database of theoretical models may provide statistical relationships among observables (intensities of [CII], [NII], CO 1$-$0, etc.) and physical quantities (gas temperature, gas column density, ISM phases, etc.). Deducing physical quantities from observables has been one of the big challenges in astrophysics since it always involves extensive modeling. Our algorithm provides an efficient way to evaluate the physical quantities from observables without extensive modeling as long as there is a large database for analysis and training that depicts statistical properties.  

\section{Acknowledgement}
The research was conducted at the Jet Propulsion Laboratory, California Institute of Technology, under a contract with the National Aeronautics and Space Administration. Preprint of an article published in https://doi.org/10.1142/S2251171723500071 \copyright World Scientific Publishing Company, https://www.worldscientific.com/worldscinet/jai

\bibliographystyle{ws-jai}
\bibliography{cii_pred}

\newpage

\appendix{Examples of Signal Fitting and Reducing Noise of SOFIA Data}

We use a signal fitter to reduce the noise of SOFIA [CII] data. The signal fitter uses an autoencoder algorithm of the convolutional neural network (CNN) \citep[see][for details]{vincent08,ying2017,hsin2019}. An autoencoder has two steps; encoder and decoder. The encoder down-samples the data while filtering unwanted features, and the decoder reconstructs the data matching the original dimension of input from the down-sampled data made by the encoder. Our signal fitter is trained to filter out high-frequency noise, mid-frequency fringes (non-Gaussian noise), and baseline fluctuations. The training data is generated by adopting multiple skewed Gaussian signals with randomized parameters for skewed Gaussian profiles and randomly selecting the number of signals between one to five. Multiple Gaussian white noises, baseline fluctuations, and mid-frequency fringes are added to signals to depict the various noise statistics. The training was done using half a million simulated spectra. The validation was made using another 0.1 million simulated spectra. The validation results demonstrated that 99\% of signals with the signal-to-noise (SNR) $>$ 3 are fitted within 1-$\sigma$ uncertainty, and 75\% of signals with the SNR of 2 $-$ 3 is fitted within 1-$\sigma$ uncertainty.

\begin{figure*}[tb]
\centering
\includegraphics[angle=0,scale=0.48]{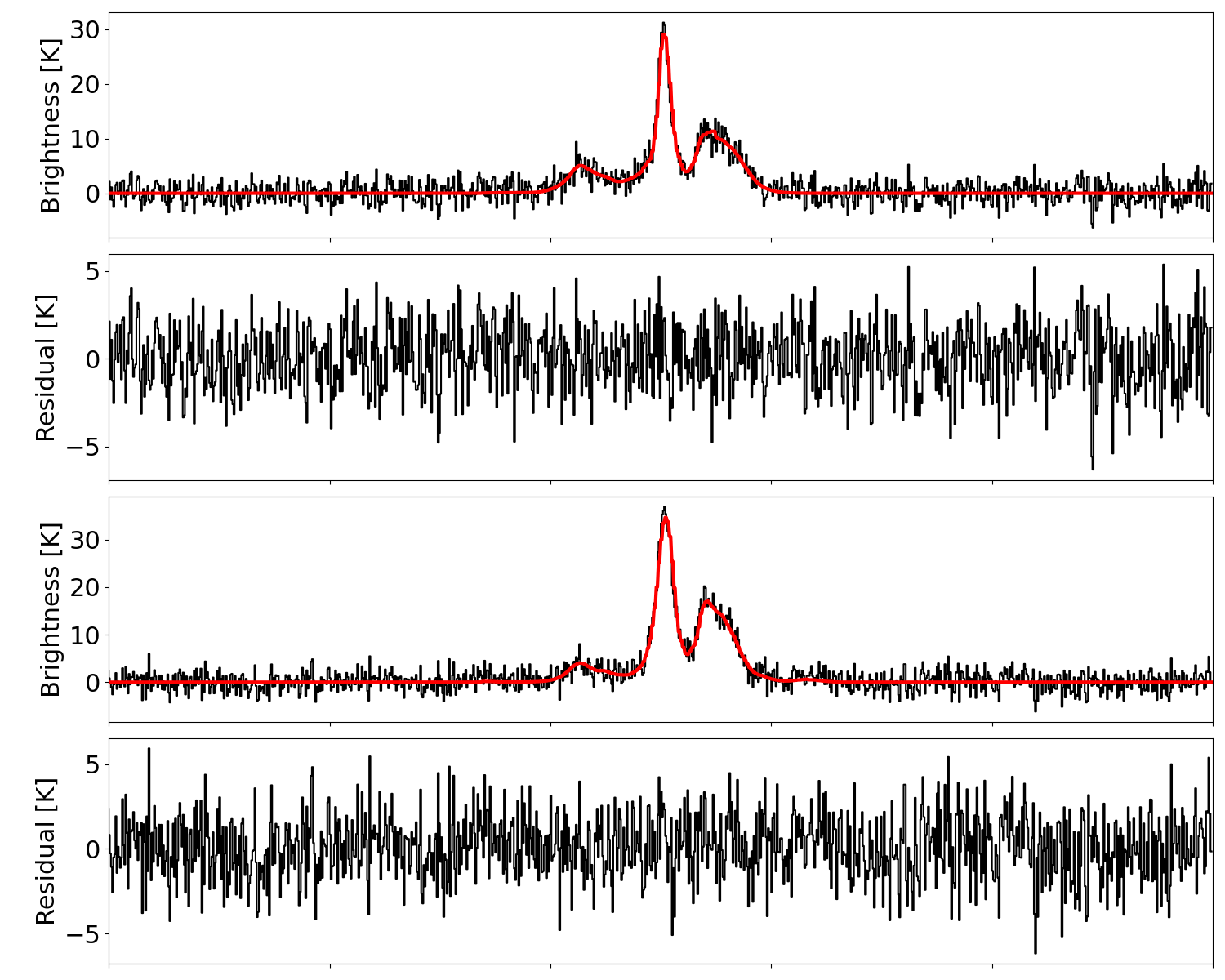}
\caption{Signal Fitting using a convolutional neural network (the first and third panels) and the residuals (the second and fourth panels). The black lines are the [CII] spectra in M17. The solid red lines are the signal fitted by a convolutional neural network.}
\label{f6}
\end{figure*}

\begin{figure*}[tb]
\centering
\includegraphics[angle=0,scale=1.1]{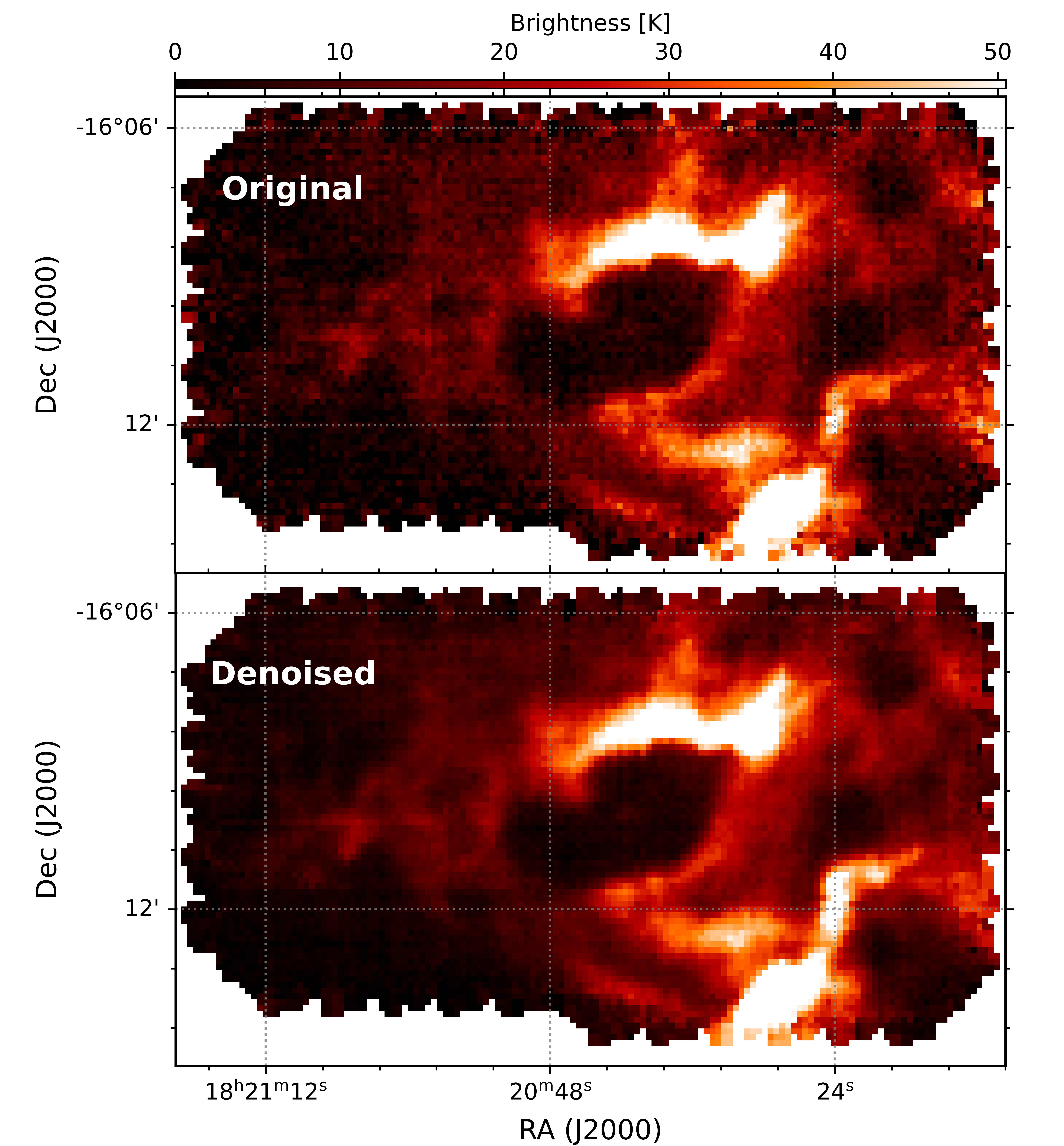}
\caption{A [CII] image of M17 at the radial velocity of 24 km/s. The top panel shows the original [CII] image. The bottom panel shows the result after noises are eliminated by the signal fitting in the spectrum at each pixel, as shown in Figure \ref{f6}. No noise reduction in the image space is made, but the signal fitting in the frequency space reduces noise in the image space as well. The noise in the image has been reduced by 50\% at a minimum throughout the image.} 
\label{f7}
\end{figure*}

Figure \ref{f6} shows the signals (solid red lines) fitted using our algorithm. The two spectra (the first and third rows) are selected from the SOFIA [CII] survey of the M17 high-mass star-forming region. The second and fourth rows show the residuals between the spectra and CNN fits in the first and third rows, respectively. The fitted signals demonstrate the successful removal of noise. We use our signal fitter rather than conventional Gaussian fitters since this algorithm is significantly faster. Conventional Gaussian fitters are often used to evaluate the physical properties of [CII] emissions. However, the main goal of the fitting signal in this study is to reduce the noise rather than evaluate the physical quantities; thus, we use our algorithms for an efficient process. Figure \ref{f7} compares the images before and after the signal fitting in the RA and Dec space (the first and second dimensions of the SOFIA data). The signal fitting is along the frequency space, which is the third dimension of the SOFIA data. While we only reduced noise in the frequency dimension, the image noise is substantially reduced (at a minimum of 50\%).

\end{document}